%% file: bare_conf.tex
\documentclass[conference]{IEEEtran}

\usepackage{multicol}

\usepackage{mathtools}

\usepackage{braket}
\usepackage{cite}
\usepackage{amsmath}
\usepackage{amssymb}
\usepackage{graphics}
\usepackage{color}
\usepackage[hang,footnotesize,bf,sf]{subfigure}
\usepackage{caption}
\usepackage{verbatim}
\usepackage{pst-all}  % pstricks
\usepackage{pst-plot}
\usepackage{color}
\usepackage{tabularx}
\usepackage{multirow}
\usepackage{booktabs}
\usepackage{threeparttable}
\usepackage{amsfonts}
\usepackage[T1]{fontenc}
\usepackage{libertine}

\DeclareGraphicsExtensions{.eps,.pdf,.png,.jpg}
\usepackage{listings}
\usepackage[USenglish]{babel}
\usepackage{fancybox}
\usepackage{multicol}
\usepackage{listings}
\usepackage[font=small]{caption}

    {\end{minipage}\end{Sbox}\fbox{\TheSbox}}
\usepackage{algorithmic} % a style for pseudocode
\usepackage{epstopdf}
\usepackage{array}
\usepackage{pbox}
\usepackage{makecell}
\usepackage{subfigure}
\newcolumntype{M}[1]{>{\raggedright\arraybackslash}m{#1}}

\DeclareGraphicsExtensions{.eps,.pdf,.png,.jpg}

\usepackage{fancybox}

\usepackage{amsmath, amssymb}

\usepackage{todonotes}

\ifCLASSINFOpdf
\else
\fi
\usepackage{hyperref}
\hypersetup{
  colorlinks   = true, %Colours links instead of ugly boxes
  urlcolor     = blue, %Colour for external hyperlinks
  linkcolor    = blue, %Colour of internal links
  citecolor   = blue%Colour of citations
}
\usepackage{url}

\usepackage[utf8]{inputenc}
\usepackage{tikz}
\usepackage{rotating}
\usepackage{amsmath,amssymb,amsthm}
\usepackage{listings}
\usepackage{wasysym}
\usepackage{pifont}
\usepackage{csvsimple}
\usepackage{siunitx}
\usepackage{tabularx}
\usepackage{xifthen}

\usepackage[T1]{fontenc}
\usepackage{dblfloatfix}
\usepackage{url}
\usepackage{subfigure}
\usepackage{qcircuit}
\usepackage{stmaryrd}
\usetikzlibrary{arrows}
\usetikzlibrary{arrows.meta}
\usetikzlibrary{positioning}  
\usetikzlibrary{shadows}
\usetikzlibrary{backgrounds,calc,chains,
	matrix,positioning,shapes,shapes.geometric,decorations,
	shapes.arrows, decorations.pathmorphing, decorations.pathreplacing, decorations.markings}

\tikzset{%
	font={\footnotesize},
	vertex/.style={draw,circle,inner sep=0pt,minimum width=0.5cm,minimum height=0.5cm},
	zeroterm/.style={below,inner sep=0pt,font=\tiny}
}

\begin{document}

\title{Full-stack quantum computing systems in the NISQ era: algorithm-driven and hardware-aware compilation techniques}

\author{\rmfamily \large Medina Bandic\textsuperscript{$\ast$}, Sebastian Feld\textsuperscript{$\ast$} and Carmen G. Almudever\textsuperscript{$\dagger$}
\vspace{0.2cm}\\
\textsuperscript{$\ast$} \small{Department of Quantum and Computer Engineering and QuTech, Delft University of Technology}\\
%\textsuperscript{$\dagger$} \small{Department of Quantum and Computer Engineering, Delft University of Technology}\\
\textsuperscript{$\dagger$} Technical University of Valencia\\
 
}

% make the title area
\maketitle

\begin{abstract}

%\textcolor{red}{Nota 1: He quitado la palabra benchmarking del título, ¿quitamos también application-driven?}

The progress in developing quantum hardware with functional quantum processors integrating tens of noisy qubits, together with the availability of near-term quantum algorithms has led to the release of the first quantum computers. These quantum computing systems already integrate different software and hardware components of the so-called  "full-stack", bridging quantum applications to quantum devices. In this paper, we will provide an overview on current full-stack quantum computing systems. We will emphasize the need for tight co-design among adjacent  layers as well as vertical cross-layer design to extract the most from noisy intermediate-scale quantum (NISQ) processors which are both error-prone and severely constrained in resources. As an example of co-design, we will focus on the development of hardware-aware and algorithm-driven compilation techniques.

\end{abstract}

\IEEEpeerreviewmaketitle

\input{intro}
\input{full-stack}

\input{mapping}

\input{Profiling}

\input{Conclusions}

% use section* for acknowledgment
\section*{Acknowledgment}
The authors sincerely appreciate scientific discussions with Prof. Eduard Alarcon (UPC).

% references section

% can use a bibliography generated by BibTeX as a .bbl file
% BibTeX documentation can be easily obtained at:
% http://mirror.ctan.org/biblio/bibtex/contrib/doc/
% The IEEEtran BibTeX style support page is at:
% http://www.michaelshell.org/tex/ieeetran/bibtex/
%\bibliographystyle{IEEEtran}
% argument is your BibTeX string definitions and bibliography database(s)
%\bibliography{IEEEabrv,../bib/paper}
%
% <OR> manually copy in the resultant .bbl file
% set second argument of \begin to the number of references
% (used to reserve space for the reference number labels box)

%\begin{thebibliography}{1}

%\bibitem{IEEEhowto:kopka}
%H.~Kopka and P.~W. Daly, \emph{A Guide to \LaTeX}, 3rd~ed.\hskip 1em plus
%  0.5em minus 0.4em\relax Harlow, England: Addison-Wesley, 1999.

%\end{thebibliography}

\bibliographystyle{unsrt}
\bibliography{sigproc}

% that's all folks
\end{document}

%% file: intro.tex
\section{Introduction}
\label{Sect1}

The general field of quantum computing has experienced remarkable progress in the last years becoming a tangible reality. Prototypes of quantum computers, also known as noisy intermediate-scale quantum (NISQ) computers \cite{preskill2018quantum}, already exist and have been made available to users through the cloud \cite{ibm17experience, qinspire}. We will still have to wait for having large-scale and fault-tolerant quantum computers that provide the expected computational power, but the potential of this new technology is undeniable \cite{moller2017impact,resch2019quantum,BGC2021}. A quantum computer will not only be capable of solving relevant problems unsolvable by current classical computers, it also represents a paradigm shift in the way of how computing is performed.  

Although there is still a long way to go and the challenges are diverse, huge advances have recently been made. Several experimental demonstrations of quantum computational advantage have been performed since 2019, when Google Research claimed to have achieved it on a 53-qubit programmable superconducting processor (Sycamore) \cite{arute2019quantum, zhong2020quantum, wu2021strong}. Furthermore, in terms of processors' scalability, qubit counts (number of qubits on a chip) are rapidly increasing, especially in quantum technologies based on superconductors. IBM just released a 127-qubit quantum processor named Eagle\cite{eagle}, and expects to present a 1000-qubit chip by 2023 \cite{IBMroadmap}. Note that adding more qubits exponentially increases the number of states the quantum computer can calculate with and thus its computational power.

\begin{figure}[t]
	\centering
	\includegraphics[width=4cm]{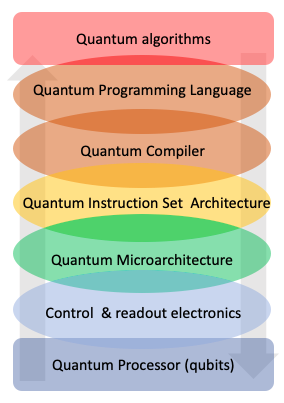}
 	\caption{Software and hardware functional elements of the quantum computing full-stack. Grey arrows represent the flow of information between hardware and software layers needed for co-design.}
 	\label{fig:stack}
\end{figure}

The progress in quantum hardware has been accompanied by advances not only on the algorithm side in the form of hybrid quantum-classical algorithms \cite{bharti2021noisy} for NISQ devices, but also on other required intermediate functionalities such as quantum software (i.e. programming languages and compilers) \cite{chong2017programming}, instruction set architecture and microarchitecture \cite{fu2019eqasm,zou2020enhancing, fu2017experimental,zhang2021exploiting} and control electronics \cite{xue2021cmos}. This has lead to the development of quantum computers, as we know them today, that integrate different software and hardware components of the so-called "full-stack" \cite{almudever2017engineering, martonosi2019next} (see also Fig \ref{fig:stack}). More precisely, full-stack quantum computing systems consist of a series of functional elements (precursors of full-fledged layers) that bridge quantum algorithms with quantum devices. Following a layered-oriented approach, which resembles classical computer architectures with some fundamental differences, quantum algorithms can be expressed using high-level programming languages and compiled to low-level instructions (e.g., quantum assembly language-instructions, QASM) that are further translated into specific signals for controlling and operating the physical qubits.

In classical computing stacks, abstractions have been introduced in the form of well-defined and self-contained layers with clear functionality that encapsulate specific information, which is only shared between adjacent layers. The level of abstraction (i.e., storing/hiding information that renders more independent layers) between different layers of the stack has been increased by virtue of the abundance of resources. This ever increasing trend resulted in software being fully independent of the underlying hardware, a desired attribute of a computer. Note though, that with the recent advent of low-power and AI co-processors, the strictly layered approach has been revisited and vertical cross-layer design and co-design are indispensable due to the scarcity of resources. 

Quantum computing systems are not ready yet for such complete abstractions due to the immaturity of its hardware. To extract the most from of NISQ processors, which are severely constrained in resources (e.g. number of qubits or qubit count) and highly prone to errors, their low-level physical details need to be exposed to higher layers of the stack. This results, for instance, in compilation techniques that consider qubits' connectivity, gate error rates, error variability across the quantum device, primitive quantum gates, and crosstalk, amongst others, to efficiently execute a quantum algorithm and increase its success rate. Thus, there is a flow of information that includes relevant hardware parameters piercing bottom-up through the stack. In addition, the compiler can leverage its knowledge about the application to perform some general (e.g. gate cancellation) or even application-specific optimization on the quantum circuit. Note that the compiler's effectiveness/efficiency is key in successfully executing a quantum algorithm and, in the long term, in the adoption of this emerging technology. 

Therefore, this early stage of quantum computing, in which resources are scarce, constrained and noisy, calls more than ever for a tight co-design among adjacent layers as well as vertical cross-layer design and related optimization of the full-stack \cite{almudever2021structured,tomesh2021quantum,shi2020,li2021software}. Note that this co-design should occur upfront, at the conception phase of the full system. The two-fold benefit of this approach would be: i) to extract the maximum computational capability out of the constrained quantum system, but also to ii) set the precursor basis of future front-ends that will pave the way towards more layer-oriented encapsulation and abstraction.

This paper will provide an overview on the full-stack quantum computing systems in the NISQ era. To this end, we will first present the current state of quantum computers showing what functional hardware and software layers they consist of and how they bridge the gap between quantum algorithms and quantum chips. Then, we will discuss about the organization or architecture of quantum computing systems, which is still far from the layered approach used in classical computer architectures. We will emphasize that, in this early-days of quantum computing, allowing information flowing across the stack is not only a must, but also brings significant gains. Hence, a discussion on the need of co-design for optimally constructing full-stack quantum computing systems will follow. We will finally provide an example on such co-design by showing how quantum circuit mapping strategies, an integral part of the overall compilation process, can be improved and potentially increase the success rate of a quantum algorithm by being not only hardware-aware, but also application-driven.

%% file: full-stack.tex
\section{Current full-stack quantum computing systems}
\label{Sect2}

Although quantum computers were recently developed, the idea of building one is quite old, as it was proposed in the early 80's by physicist Richard Feynman \cite{feynman1982simulating} for simulating quantum systems. Ten years later the first quantum algorithms appeared \cite{bharti2021noisy}.
It was not till late 90's, when experimental demonstrations of a quantum algorithm running on a two-quit quantum processor were shown \cite{chuang1998experimental}. Since then, quantum hardware has substantially progressed, especially in the last years; a variety of technologies for qubit implementation, such as quantum dots, solid-state spins, trapped-ions or superconductors, are being explored \cite{resch2019quantum}. In addition, its main (still) limiting characteristics are constantly improved with increased qubit counts, larger coherence times and higher gate fidelities. Note that, in spite of this progress, quantum processors are still in their infancy, being extremely resource-constrained and error-prone.

At the top of the stack, quantum algorithms have also evolved following the capabilities of quantum hardware. Most of the algorithms initially proposed (e.g., Shor, HHL) cannot be executed on current nor near-future quantum machines, as they require the incorporation of quantum error correction and fault-tolerant techniques, and therefore an immense number of qubits (in the order of millions). That is why the quantum algorithm community has made an effort to develop algorithms suitable for NISQ processors \cite{bharti2021noisy}.

To close the gap between quantum algorithms and devices, and to be able to run quantum algorithms on larger and larger quantum processors, further components needed to be added and integrated to complete the system such as the software part, the interface between software and hardware, and the control stack. This led to the development of modern full-stack quantum computing systems, which were inspired by classical computer architectures, and include the following elements (see Fig. \ref{fig:stack}): quantum applications, high-level quantum programming languages and compilers, quantum instruction set architecture and related microarchitecture, control electronics and the quantum device. Although essential functional elements of the quantum computing full-stack have been identified and integrated, the architecture and organization of such stacks may differ between systems and is in constant evolution as the field progresses \cite {martonosi2019next}. 

For current quantum computers, algorithms can be programmed using programming frameworks\cite{chong2017programming, khammassi2020openql}. They not only feature high-level programming languages such as Python to describe quantum circuits, but also compilers that translate those high-level instructions into low-level ones, usually expressed in a quantum assembly-like language understandable and therefore executable by the underlying processor. As further explained in Sec. \ref{Sec3}, quantum compilers are also responsible for making transformations to the quantum circuit to fulfill the quantum hardware's constraints, along with optimizations to reduce the circuit depth, for instance. Note that quantum compilers are located in the middle of the stack, as they are key to bridge quantum applications to quantum devices. The output of the compiler, low-level instructions, are then further translated into specific pulses to operate and control the chip's qubits. For this purpose, quantum instruction set architectures\cite{fu2019eqasm,zou2020enhancing} and microarchitectures \cite{fu2017experimental,zhang2021exploiting} as well as control electronics (at even cryogenic temperatures) \cite{xue2021cmos} have been developed.

\subsection*{Towards increased abstraction in quantum computing systems: the need for co-design to architect}

Classical computers consist, from an architectural point of view, of a series of layers with well-defined functionality in which relevant information is encapsulated and only exposed to adjacent levels. This layered architecture has evolved with a growing level of abstraction between layers as the amount of resources has been substantially increased.   

Most of the functional elements present in classical systems, already in the form of layers, can also be identified in current quantum computers, but not as mature though. In full-stack quantum computing systems, there is still a tight interplay among and across functionalities and therefore no strict layered abstracted architecture yet with clear allocation of functionality per layer exists. This is mainly due to severely limited and impaired resources in quantum devices, which encompass: limited qubit count and connectivity among them, short coherence time, high operation error rates, and crosstalk. Actually, allowing information from the physical lowest-level (i.e. quantum processor) to flow up to high-level functions is not only a must for being able to (successfully) execute an algorithm, but the best way to maximize performance \cite{shi2020}. A notable example of such information flow is within the compiler, which uses quantum processor characteristics to modify the circuit for meeting the hardware's constraints but also to maximize the corresponding success rate. Instances of that include the introduction of software techniques to deal with or alleviate crosstalk \cite{murali2020software, ding2020systematic} and noise-aware compilation methods \cite{murali2019full}. Note that quantum hardware information can be combined with algorithm parameters to further optimize compilation techniques given a specific application (application-specific compilers) \cite{lao20212qan,li2021software}.

These examples, in which information exchange is bounded, punctual and limited, constitute ad-hoc predecessors of full co-design techniques. As defined in \cite{tomesh2021quantum}, co-design refers to "the flow of information between different hardware and software stack layers, in order to improve the overall application execution and hardware design. The information flow might include: key hardware parameters, design specifications, and resource requirements up and down the stack. Co-design for quantum computing is about incorporating this information into the techniques and system designs at every layer of the stack to make optimal use of limited resources''. Culminating co-design, both across adjacent layers as well as cross-layer vertical design, will require these techniques to be all-pervasive in coverage, information-rich in exchange, and structured in their application \cite{almudever2021structured}. Finally, we postulate that co-design is not an aim per se, but a means to  eventually achieve full abstraction, since this information exchange across layers will serve as a basis to implement front-ends that allow self-contained encapsulated stack layers.

The next section will provide a specific example of quantum co-design. More precisely, it will show how the mapping process, which is an essential part of the compiler, can be further optimised by considering not only hardware characteristics, but also relevant algorithm properties.

%% file: mapping.tex
\section{Quantum circuit mapping: Accommodating quantum algorithms to resource-constrained quantum devices }
\label{Sec3}

As discussed above, quantum computing full-stacks have been developed to enable the execution of quantum algorithms on quantum processors. Current NISQ devices can only handle simpler algorithms, in terms of number of gates and circuit depth, as they are still constrained by the presence of noise: quantum gates have high error rates and qubits are fragile and decohere over time resulting in information loss. On top of that, running an algorithm on a NISQ device is not a straightforward process, as there are other hardware limitations that must be considered.

\begin{figure}[t!]
	\centering
	\includegraphics[width=9cm]{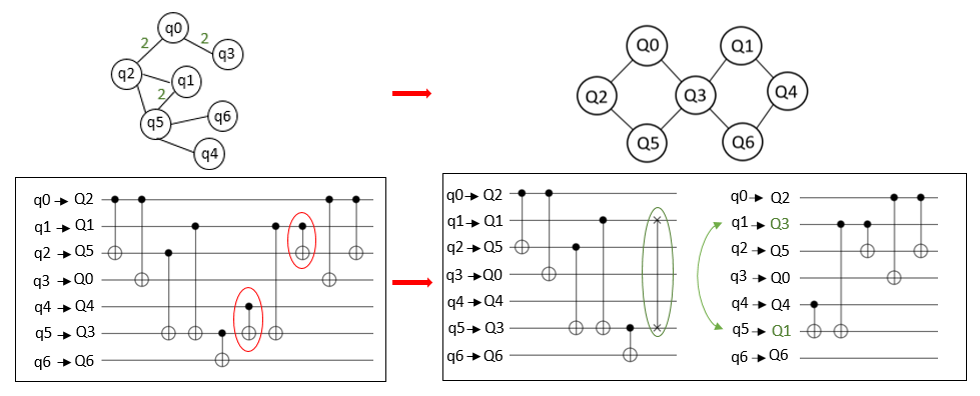}
 	\caption{Running a quantum circuit on a Surface-7 quantum processor \cite{versluis2017scalable}. Top-left: Interaction graph of the circuit shown below. 
 	  Top-right: The chip coupling graph; nodes represent physical qubits and edges show connections on the chip (possible interactions). Bottom: Qubits in the circuit ($qi$) are mapped onto physical qubits labeled with $Qi$. An extra SWAP gate is required for being able to perform all CNOT gates. \cite{bandic2020structured}}
 	\label{fig:mapping1}
\end{figure}

One of the most relevant quantum hardware constraint is  \emph{limited qubit connectivity}. For most technologies, including superconducting qubits and quantum dots, qubits are arranged in a 2D grid topology (see Fig. \ref{fig:mapping1}, top right) allowing only nearest-neighbor interactions. This means that in order to perform a two-qubit gate, the two interacting qubits have to be placed in neighboring locations on the chip, which is not always possible (Fig. \ref{fig:mapping1}). Other constraints that affect algorithm execution are: i) \emph{primitive gate set}, that is, the quantum gates supported by the device. Note that a quantum chip gate set does not necessarily have to match the one used in the circuit to be run; and ii) \emph{classical control constraints} that come from the use of shared control electronics required to scale up quantum computing systems. This limits the operations' parallelization. It is the responsibility of the compiler, and more precisely of \emph{the mapping process}, to satisfy all device constraints while accommodating the algorithm's needs.
 
\begin{figure*}[h!]
    \centering
    \centerline{
\subfigure[]{\includegraphics[width=2.5in]{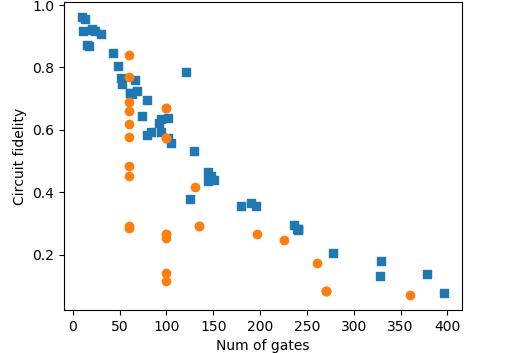}\hspace{-6mm}%
\label{fig:fidDecGates}}
\subfigure[]{\includegraphics[width=2.5in]{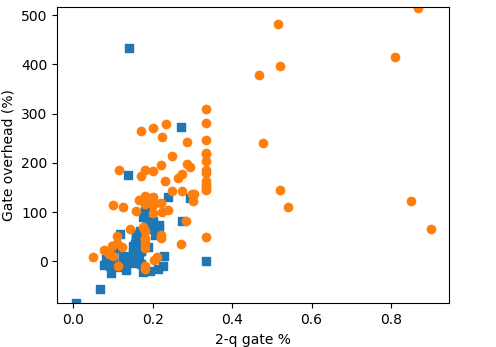}\hspace{-6mm}%
\label{fig:gateOv2q}}
\subfigure[]{\includegraphics[width=2.5in]{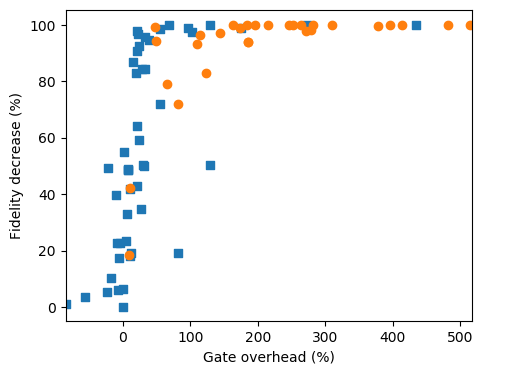}
\label{fig:fidDec}}}
    \caption{Impact of the circuit mapping process. (a) Gate number vs. circuit fidelity, (b) 2-qubit gate \% influence on gate overhead, and (\%) (c) Gate overhead and decrease in fidelity relation, when randomly generated circuits (blue squares) and real algorithms (orange circles) taken from \cite{qbench} are mapped into an extended 100-qubit version of the Surface-17 hardware configuration. To this purpose, the trivial mapper integrated in the OpenQL compiler \cite{khammassi2020openql} has been used. For a) and c) only circuits with less than 400 gates were used. Circuit fidelity is calculated as product of fidelities for all one- and two- qubit gates in the circuit, based on the error-rate values taken from \cite{versluis2017scalable}.
    } 
\label{fig:graphs1}
\end{figure*}

The quantum mapping process consists of the following steps (not necessarily in this particular order), see also Fig. \ref{fig:mapping1}: 1) \emph{Decomposition of the gates} of the circuit to the primitive gate set; 2) \emph{Scheduling} quantum operations to leverage parallelism and therefore shorten execution time; 3) Smartly \emph{placing virtual qubits} (from the circuit) \emph{onto physical qubits} (placements on actual chip) such that the previously mentioned nearest-neighbor two-qubit gate constraint is satisfied as much as possible during circuit execution; and 4) \emph{Routing} or exchanging positions of virtual qubits on the chip such that all qubits that need to interact during circuit execution are adjacent. This is done by inserting additional quantum gates called SWAPs, whose purpose is to exchange virtual qubit values (i.e., quantum states) between physical qubits on the chip. Therefore, the resulting circuit after-mapping might have more gates and higher depth, and hence a longer execution time than the original one. Due to the previously mentioned highly-erroneous quantum operations and qubit decoherence, it is crucial to create only minimal overhead as it affects the algorithm's fidelity: as the number of gates in the circuit increases, the fidelity decreases (see Fig. \ref{fig:fidDecGates}).

There exist various approaches to solve the mapping problem, each using different methods and strategies \cite{almudeverrealizing, lao2021timing, pozzi2020using, hillmich2021exploiting, itoko2020optimization,tan2021optimal,jiang2021quantum,li2020qubit}. 
They also differ in the cost function optimized in the mapping process, which is also used as the performance metric to assess the mapper. Usual metrics are gate overhead (number of SWAPs), circuit depth and latency overhead (number of time-stamps) and reliability/fidelity or success rate probability. Fig. \ref{fig:gateOv2q} shows how the percentage of two-qubit gates in the circuit affects the gate overhead: the higher this percentage, the more qubits interact and therefore the higher the gate overhead caused by routing. Additional (SWAP) gates then impact the circuit's fidelity as shown in Fig. \ref{fig:fidDec}. 

The majority of circuit mapping approaches mostly focus on device characteristics without considering structural properties of the quantum circuit itself. When characterizing benchmark circuits, the only parameters taken into account in literature are gate and qubit count, two-qubit gate percentage (Fig. \ref{fig:graphs1}) and at times circuit depth. More in-depth algorithm characterization is of crucial importance because it enable enable us to: i) perform an exhaustive comparison and taxonomy of algorithms; ii) perform an in-depth analysis on the performance of compilation (mapping) techniques, and iii) further improve the compilation process creating not only hardware-aware but also algorithm-driven solutions.

\begin{figure}[b]
	\centering
	\includegraphics[width=7cm]{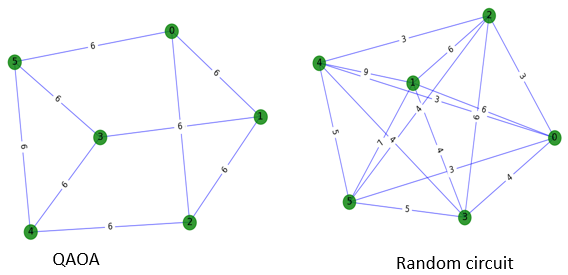}
 	\caption{Interaction graphs of circuits with the same size parameters: num. of qubits = 6, num. of gates = 456, 2-qubit gate \% = 0.135.}
 	\label{fig:interaction}
\end{figure}

Some works have already pointed out the importance of considering algorithm properties \cite{lubinski2021application,mills2020application,li2020towards} for further improving the compilation techniques and, in particular, interaction graphs for the mapping procedure \cite{steinberg2021noise,bandic2020structured}.
\emph{Interaction graphs} are graphical representations of the two-qubit gates of a given quantum circuit. Fig. \ref{fig:mapping1} shows an example of a quantum circuit along with its interaction graph representation. Edges represent two-qubit gates and nodes are the qubits that participate in those. If a circuit comprises multiple two-qubit gates between pairs of qubits, it results in a weighted graph (like in Fig. \ref{fig:mapping1}) which shows how often each pair of qubits interacts and how those interactions are distributed. This additional information can be leveraged to provide more insight into the structure of a circuit that is otherwise hidden when only considering common algorithm parameters such as number of qubits and gates and two-qubit gate percentage. To illustrate this, Fig. \ref{fig:interaction} shows the interaction graphs of two quantum algorithms, a real one (QAOA, on the left) and a randomly generated circuit (on the right), with the same properties when only characterized in terms of the three common algorithm parameters. What can be noticed is that their interaction graph structure is quite different: the graph of the random circuit is more complex with full-connectivity and present a different distribution of the interactions between qubits, that is, of the weights. This will result in more routing and therefore higher overhead. As shown in Fig. \ref{fig:graphs1}, the gate overhead and fidelity decrease is, on average, higher for synthetic (random) algorithms than for the real ones.

\newpage

Therefore, analysing interaction graphs might help us understand why a mapping solution works better for specific (groups) of algorithms first, and then come up with optimised mapping techniques that are both, algorithm-driven and hardware-aware. Algorithm-driven devices could be an effective solution in dealing with limited NISQ computing resources \cite{li2020towards,tomesh2021quantum}, as they can precisely be designed for some dedicated purpose. Even in classical computing used in daily lives, different computing assets are required for different specific purposes (applications).

%% file: profiling.tex
\section{Algorithm-driven mapping solutions: characterising qubit interaction graphs}
\label{Sec4}

As explained in the previous section, we will broaden the scope of algorithm characterization by introducing interaction-graph-based profiling. 
We will discuss how graph parameters might be meaningful for improving quantum circuit mapping techniques. Some  preliminary results will be presented, showing the relation of those graph parameters to the performance of circuit mapping.

The main reason and importance of exploring the properties of interaction graphs is the fact that they represent the core constraint that needs to be dealt with during the mapping process: two-qubit gates and their dispersion among pairs of qubits. For that purpose we took input from graph theory and characterized quantum algorithms based on their interaction graph metrics \cite{hernandez2011classification} such as average shortest path, connectivity, clustering coefficient and similar ones, with a focus on metrics that are of interest for the mapping problem.

\begin{table}[b!]
	\centering
	\includegraphics[width=9cm]{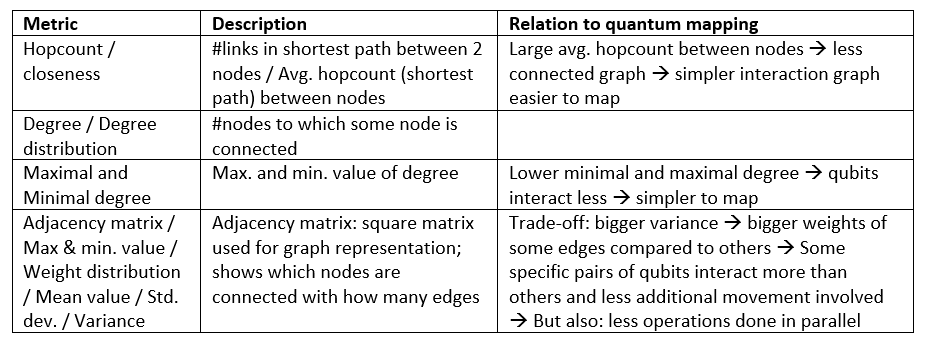}
 	\caption{Metrics for characterizing interaction graphs \cite{hernandez2011classification} and their relation to mapping.}
 	\label{tbl:metr1}
\end{table}

What can be noticed is that large number of handpicked, mapping-related metrics is codependent, i.e. they scale in the same manner. In order to reduce the parameter space and select only features that are necessary, a Pearson correlation matrix was created. Applying this method reduced our previous metric set to: average shortest path (hopcount/clossenes), maximal and minimal degree and adjacency matrix standard deviation, as shown in Tab. \ref{tbl:metr1}. Using this new metrics and the common circuit parameters, algorithms can be clustered based on their similarities. Ideally, quantum algorithms with similar properties are ought to show similar performance when run on specific chips using a given mapping strategy.

\begin{figure}[h]
    \centering
    \includegraphics[width=4in]{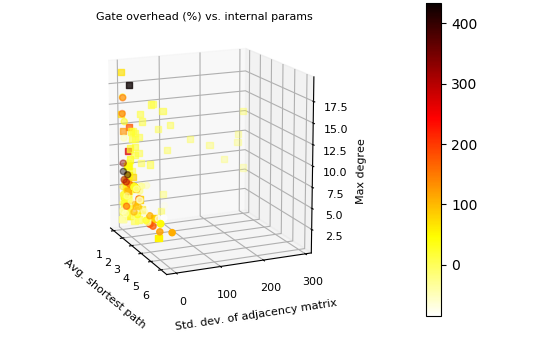}
    \caption{Gate overhead (\%) vs. inter. graph parameters. Each point represents a benchmark mapped on the chip. Synthetically generated circuits are marked with a square, real ones with a circle.}
    \label{fig:gateInt}
\end{figure}

Some preliminary results on the use of graph-based metrics, with the final aim to further optimise the quantum circuit mapping process, are shown in Fig. \ref{fig:gateInt}. We have compiled 200 quantum circuits by using the same hardware and mapping configuration as described in caption of Fig. \ref{fig:graphs1}. The benchmark set used \cite{qbench} contains circuits of a large variety in size (1-54 qubits, 5-100000 gates, 10-90\% two-qubit gate percentage) and type (random, reversible ones \cite{wille2008revlib} and those corresponding to real algorithms). Our first goal was to analyse how the graph-based metrics relate to gate overhead and consequently algorithm fidelity decrease.

Fig. \ref{fig:gateInt} shows that all circuits with high gate overhead had on average low variation in edge weight distribution, low average shortest path between qubits and higher max. degree, which are expected values from Tab. \ref{tbl:metr1}.

%% file: Conclusions.tex
\section{Conclusions}

Current intermediate-scale quantum computers already integrate the different hardware and software elements of the so-called full-stack, allowing the execution of quantum algorithms on NISQ devices. In this early-stage of quantum computing, quantum processors are resource constrained and highly error-prone, which is preventing these stacks from following an abstracted layered approach as used in classical computers. In terms of computer architecture,  there  is  still  a  tight interplay among and across layers, which shows a progressive and  continuously changing allocation of functionalities and requires the addition of new ones as the field advances. In other words, the organization of the stack is in constant evolution and it is being shaped based on the quantum hardware capabilities as well as on the quantum algorithms requirements. A higher level of abstraction between layers is expected as the technology matures.

A crucial aspect for architecting not only nowadays but also near-term quantum computing full-stacks, which are expected to be in the form of application-specific quantum accelerators, is co-design. Allowing information flowing up and down across the stack is key to optimise the stack and extract the maximum computational power out of the constrained quantum system. Although co-design techniques are already being used in the development of, for instance, quantum compilers or even quantum algorithms, applying co-design in its broadest sense, will require these techniques to be all-pervasive in coverage, information-rich in exchange, and structured in their application. We postulate that co-design will be also a mean to eventually achieve full abstraction, since this vertical information exchange across the stack will help to implement front-ends allowing the development of self-contained encapsulated layers.

Finally, quantum compilers, and more precisely the mapping of quantum circuits, are the most sensitive parts of the stack for co-design as their main functionality is to make some transformations to the quantum circuits that they can be efficiently run on a given quantum device. To further improve the compilation techniques is key for achieving higher algorithm success rates in the NISQ era and therefore, towards showing "quantum practicality or usefullness''. The optimization of the mapping methodologies, requires them to be not only hardware-aware but also algorithm-driven. This calls for a more in-depth profiling of quantum applications. We show instances that looking into the qubit interaction graph and considering graph-based metrics might assist, guide, dimension and optimize such mapping techniques.